\documentclass[aps,prd,showpacs,amsmath,nofootinbib,superscriptaddress,twocolumn,
preprintnumbers]{revtex4-1}
\usepackage{epsfig,amsmath,subfigure}
\usepackage{blindtext}
\def\gr{{$\gamma$-ray\,}}

\def\bs{\bigskip}

\def\ea{\ et al. \,}

\def\be{\begin{equation}}
\def\ee{\end{equation}}

\def\g{\gamma}
\def\gt{\gamma_t}
\def\gti{\gamma_{t_i}}
\def\gte{\gamma_{t_e}}

\def\gee{\gamma_e}
\def\gp{\gamma_p}
\def\rd{r_d}
\def\rde{r_{d_e}}

\def\ri{\overrightarrow{r_i}}
\def\ro{\overrightarrow{r_o}}
\def\rin{r_i}
\def\ron{r_o}
\def\zi{t_i}
\def\zo{t_o}
\def\tm{t_m}
\def\rmm{r_m}
\def\to{\theta_o}
\def\ti{\theta_i}
\def\fho{\phi_o}
\def\fhi{\phi_i}
\def\bs{b_*}

\def\sinel{\sigma_{pp}^{in}}
\def\sppi{\sigma_{pp}^{\pi}}

\begin{document}

\title{Galactic Energetic Particles and Their Radiative Yields in Clusters}

\author{Yoel Rephaeli}

\altaffiliation[Also at ]{Center for Astrophysics and Space Sciences, University of California, 
San Diego, La Jolla, CA 92093-0424}

\author{Sharon Sadeh}

\affiliation{School of Physics and Astronomy, Tel Aviv University, Tel Aviv, 69978, Israel}
\email{yoelr@wise.tau.ac.il}

\date{\today}

\begin{abstract}
As energetic particles diffuse out of radio and star-forming galaxies (SFGs), their intracluster 
density builds up to a level that could account for a substantial part or all the emission from a 
radio halo. We calculate the particle time-dependent, spectro-spatial distributions from a solution 
of a diffusion equation with radio galaxies as sources of electrons, and SFGs as sources of both 
electrons and protons. Whereas strong radio galaxies are typically found in the cluster (e.g., Coma) 
core, the fraction of SFGs increases with distance from the cluster center. Scaling particle escape 
rates from their sources to the reasonably well determined Galactic rates, and for realistic 
gas density and magnetic field spatial profiles, we find that predicted spectra and spatial profiles 
of radio emission from primary and secondary electrons are roughly consistent with those deduced 
from current measurements of the Coma halo (after subtraction of emission from the relic Coma A). 
Nonthermal X-ray emission is predicted to be mostly by Compton scattering of electrons from 
radio galaxies off the CMB, whereas $\gamma$-ray emission is primarily from the decay of neutral 
pions produced in interactions of protons from SFGs with protons in intracluster gas. 
\end{abstract}

\maketitle

\section{Introduction}

A quantitative study of energetic particles (`cosmic rays') and magnetic fields in galaxy 
clusters is motivated by the need for a more complete physical description of these systems, 
and for improved knowledge of the origin of non-thermal (NT) quantities in extragalactic 
environments. Current and future measurement capabilities necessitate fairly detailed 
modeling of energetic particles, their origin and impact on the properties of intracluster 
(IC) gas. 

Direct evidence for the presence of relativistic electrons in IC space is presently limited 
to measurements of extended regions of radio emission, `halos' and relics, which have already 
been observed in many clusters [\onlinecite{fer08}]. Mapping these extended low brightness 
regions is quite challenging due to the inherently imprecise process of subtracting out the emission 
from many galactic radio sources. Searches for NT X-ray emission have not yet yielded conclusive 
results (e.g., [\onlinecite{r08}]). Analysis 
of superposed {\it Fermi}/LAT measurements of 50 clusters 
yielded an upper limit on the mean emission above $\sim100$ MeV [\onlinecite{ack14}].
Radio measurements provide direct evidence for acceleration of electrons in SN shocks, 
whereas recent {\it Fermi}/LAT measurements [\onlinecite{ack13}] reveal the signature of 
acceleration also of protons, whose interactions with ambient protons produce neutral pions 
($\pi^{0}$) which decay into two ($\geq 70$ MeV) photons, giving rise to the `pion bump' in 
the measured spectra. Since energetic electrons and protons do not lose all their energy while 
diffusing out of their galactic sources, star-forming galaxies (SFGs) are likely to contribute 
appreciably to their respective extragalactic energy densities. While the sources of the radio 
emitting electrons have not yet been fully identified these clearly include also radio galaxies.

It has been suggested long ago that the (typically) transonic motion of galaxies through IC plasma 
excites turbulence that may re-accelerate particles, maintaining electron populations that could 
explain the observed radio halos [\onlinecite{j80}]. The main focus has been in merger and 
accretion shocks as particle acceleration sites (e.g., [\onlinecite{fts03}], [\onlinecite{bj14}]). 
Particle acceleration by mildly supersonic shocks in the dilute IC gas is not fully understood, 
and the efficiency of conversion of kinetic to particle energy is substantially uncertain. 

There is considerable motivation to model NT phenomena under typical conditions in most 
clusters. Realistic estimates of the emission from particles originating in SFGs yield lower 
limits on IC emission. Together with direct emission from the galaxies, a minimal level 
of the cluster total emission is obtained. Adding the emission from SFGs to the similarly 
deduced emission of strong radio galaxies yields the total galactic and IC emission. A basic 
issue is whether the combined emission from all galactic sources can fully account for 
emission from a cluster radio halo. This important consideration has obvious implications 
for the need to assume efficient particle re-acceleration in IC space.

The above considerations warrant a detailed assessment of galactic origin for energetic particles 
in clusters, our main objective here. In Section 2 we calculate particle spectra from SF and central 
radio galaxies, and present results of the calculations for the particle radiative spectra in Section 3. 
A brief discussion follows in Section 4.

\section{Particle Distributions}

Particles random-walk through the magnetized IC gas with their propagation likely controlled by 
magnetic field inhomogeneities, quantified by the field coherence scale denoted by $l$, and an 
implied diffusion coefficient $D=l c/3$. The particle time-dependent, spectro-spatial distribution is
governed by the spherical diffusion equation [\onlinecite{aav95}]
\begin{equation}
\frac{\partial f}{\partial t}=\frac{D}{r^2}\frac{\partial}{\partial r}
\biggl (r^2 \frac{\partial f}{\partial r}\biggr)
+\frac{\partial}{\partial\gamma}
\left( bf \right)+Q,
\end{equation}
where $f(r,t,\gamma)$ denotes the spectral particle density at time $t$, $\gamma$ is the 
particle Lorentz factor, $b(\gamma)=-d\gamma/dt$ is the energy loss rate, and 
$Q(r,t,\gamma) \propto \delta(r)\delta(t-t_0)\delta(\gamma-\gamma_0)$ is a source term 
for initial injection of monoenergetic particles at $r=0$, expressed as the product of Dirac 
delta functions in $r$, $t$, and $\gamma$. For a (functionally) separable spectral dependence 
of the source term, $q(\gamma)$, the solution can be obtained from the Green function in 
$r$ and $t$ [\onlinecite{aav95}] 
\begin{equation}
f(r,t,\gamma)=\frac{ q(\gt)b(\gamma_t)}{\pi^{3/2}b(\gamma)r_d^3}\exp\left(-\frac{r^2}
{r_d^2}\right),
\end{equation}
where $\gamma_t$ is the initial value of the Lorentz factor that reduces to $\gamma$ after a 
time $t =\int_{\gamma}^{\gamma_t}dx/b(x)$, during which the particle diffuses a distance $r_d =2\sqrt{Dt}$. 

The superposed contributions of sources at positions $\ri$ to the particle distribution function at 
time $\zo$ and observation point $\ro$ is 
\begin{equation}
\begin{split}
&f(\ro,\zo,\gamma)\approx\int_{t_m}^{\zo}
d\zi\int_{0}^R\int_0^{\pi}\int_0^{2\pi}
\frac{q(\gti)b(\gti)}{\pi^{3/2}b(\g)\rd^3}\\
&\times\exp\left(\frac{-|\ro-\ri|^2}{\rd^2}\right) 
\rin^2 h(\rin)d\rin\sin{\ti}d\ti d\fhi , 
\end{split}
\label{eq:s2c}
\end{equation}
where $h(\rin)$ is the source density
\footnote{Cluster galaxies (excluding a galaxy at the dynamical center) move substantial 
distances during characteristic times they are active sources of energetic particles; thus, 
their density can be approximated by a continuous function} 
at radial position $\rin$, and $t_m$ is the lowest injection time 
over which a particle initial energy $\gamma_{t} \rightarrow \infty$ is reduced to
$\gamma$. In spherical coordinates,
\begin{equation}
\begin{split}
&\left(\ro-\ri\right)^2=\ron^2+\rin^2 -2\ron\rin \\
&\times \left[\sin{\to}\sin{\ti}\cos{(\fho-\fhi)}
+\cos{\to}\cos{\ti}\right].
\end{split}
\label{eq:riro}
\end{equation}
Assuming a spherically symmetric source distribution, the spectral density at $\ro$ is independent 
of the angular variables; by setting $\to=\fho=0$ and integrating over $\ti$ and $\fhi$ we obtain 
\begin{equation}
\begin{split}
&f(\ron,\zo,\gee)=\frac{2\pi^{-1/2}}{b_e({\gee})}\int_{t_m}^{\zo}dt_i \int_0^R\frac{q(\gti)b(\gti)}
{\ron\rde}\rin h(\rin)\\
&\times\exp\left(-\frac{\ron^2+\rin^2}{\rde^2}\right)\sinh\left(\frac{2\ron\rin}{\rde^2}\right)d\rin.
\label{eq:fgd}
\end{split}
\end{equation}
Substituting the particle escape and energy loss rates (specified below) in the above equation 
fully determines the time-dependent, spectro-spatial distribution. We treat separately the evolution 
of electrons and protons from the very extended distribution of SFGs and that of electrons from strong 
radio sources in the cluster core.

\subsection{Electrons from Central Radio Galaxies}

In many radio halo clusters there are dominant radio galaxies within the core region, and a 
population of elliptical radio galaxies whose distribution is centrally peaked. For modeling 
simplicity, we consider the case when there are no strong radio sources outside the core 
and the total contribution of radio galaxies outside this region can be ignored. Due to 
the special location of the central (AGN-powered) radio galaxies the period of their 
activity (which is likely to be sustained by accretion and mergers), either continuously or 
intermittently, is longer than a typical lifetime of a radio galaxy. During this period radio 
sources that are not at the dynamical center move substantial distances across the cluster 
core. Therefore, we model these galaxies as a spatially uniform core component with a 
source term that is scaled to their total radio luminosity and the mean value of their magnetic 
field. 

\subsection{Electrons \& Protons from Star-Forming Galaxies}
 
The close link between high-mass stars and particle acceleration provides a basis for estimating 
the total number of Galaxy-equivalent SFGs, $N_{s}$, from the ratio of the cluster blue luminosity 
to the blue luminosity of the Galaxy. In the Coma cluster this ratio is $\sim (1-2)\times10^{2}$; we 
scale $N_{s}$ to the lower value. Due to the considerable evidence for star formation in some S0 
and elliptical galaxies (e.g., [\onlinecite{rea15}]), $N_{s}$ is not simply the number of spiral galaxies. 

The radial distribution of the fractional number of SFGs can be derived from their measured 
fraction in a compilation of 55 clusters [\onlinecite{wgj93}]. Up to a projected distance 
$R_*\sim 2$ Mpc the measured fraction is well fit by $a_{1}\bs/(a_{2}+\bs)$, where 
$(a_{1},a_{2})\sim(0.6,0.4)$, and $\bs$ is the projected distance. Assuming cylindrical 
symmetry about the los through the cluster center, the continuum limit of the SFG density, 
$h(r)$, can be calculated from an inverse Abel transform, 
\begin{equation}
h(r)=\frac{N_{s}}{4\pi}\frac{a_{3}+a_{4}r}{r^2+a_{5}r^3+a_{6}r^4}\left[\int_0^R
\frac{a_{3}+a_{4}r}{1+a_{5}r+a_{6}r^2}dr\right]^{-1},
\end{equation}
with $(a_{3},a_{4},a_{5},a_{6})\sim(-.01,.43,.69,.28)$.

\begin{figure*}[thp]
\centering
\epsfig{file=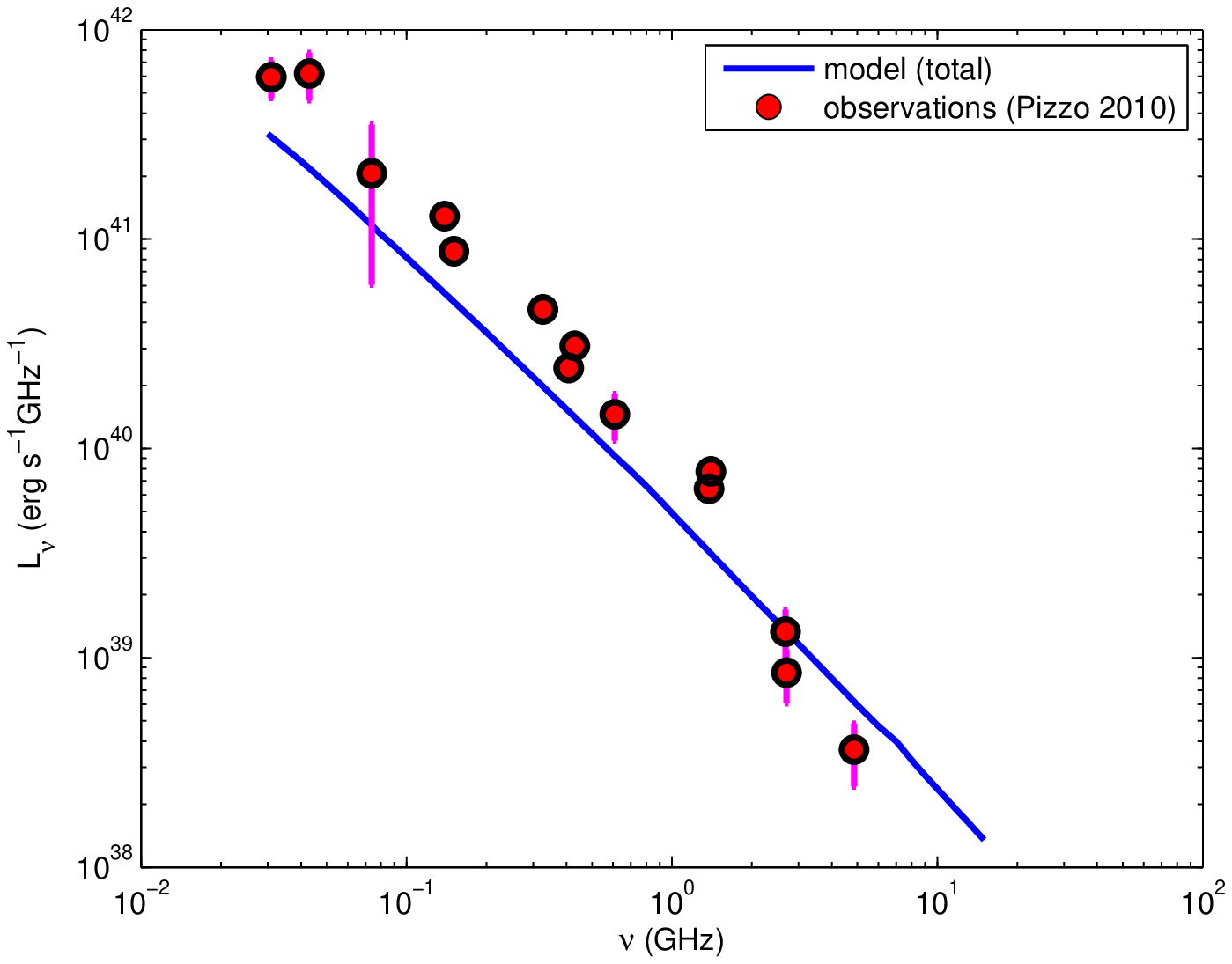,width=5.7cm,height=4.5cm}
\epsfig{file=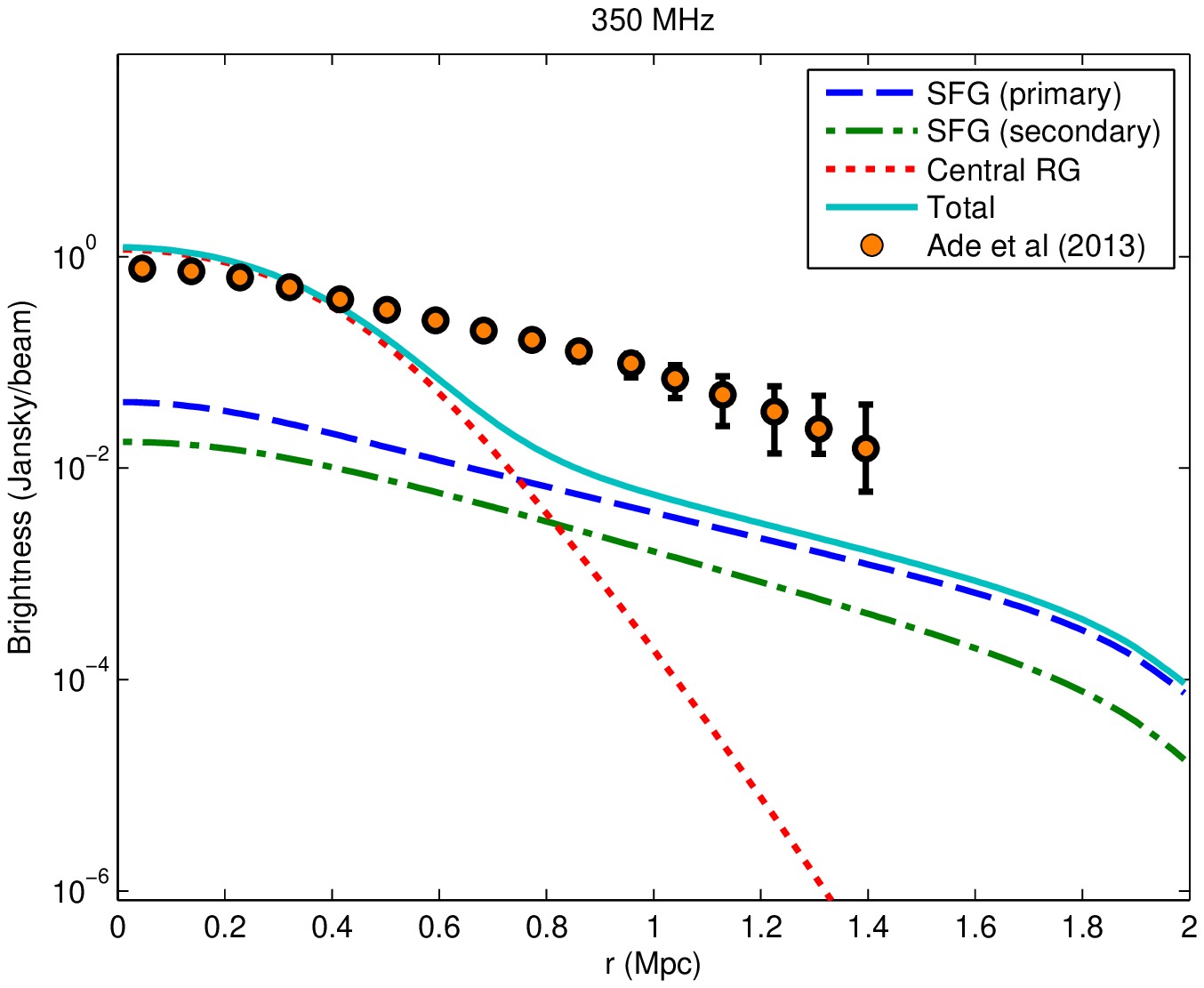,width=5.7cm,height=4.5cm}
\epsfig{file=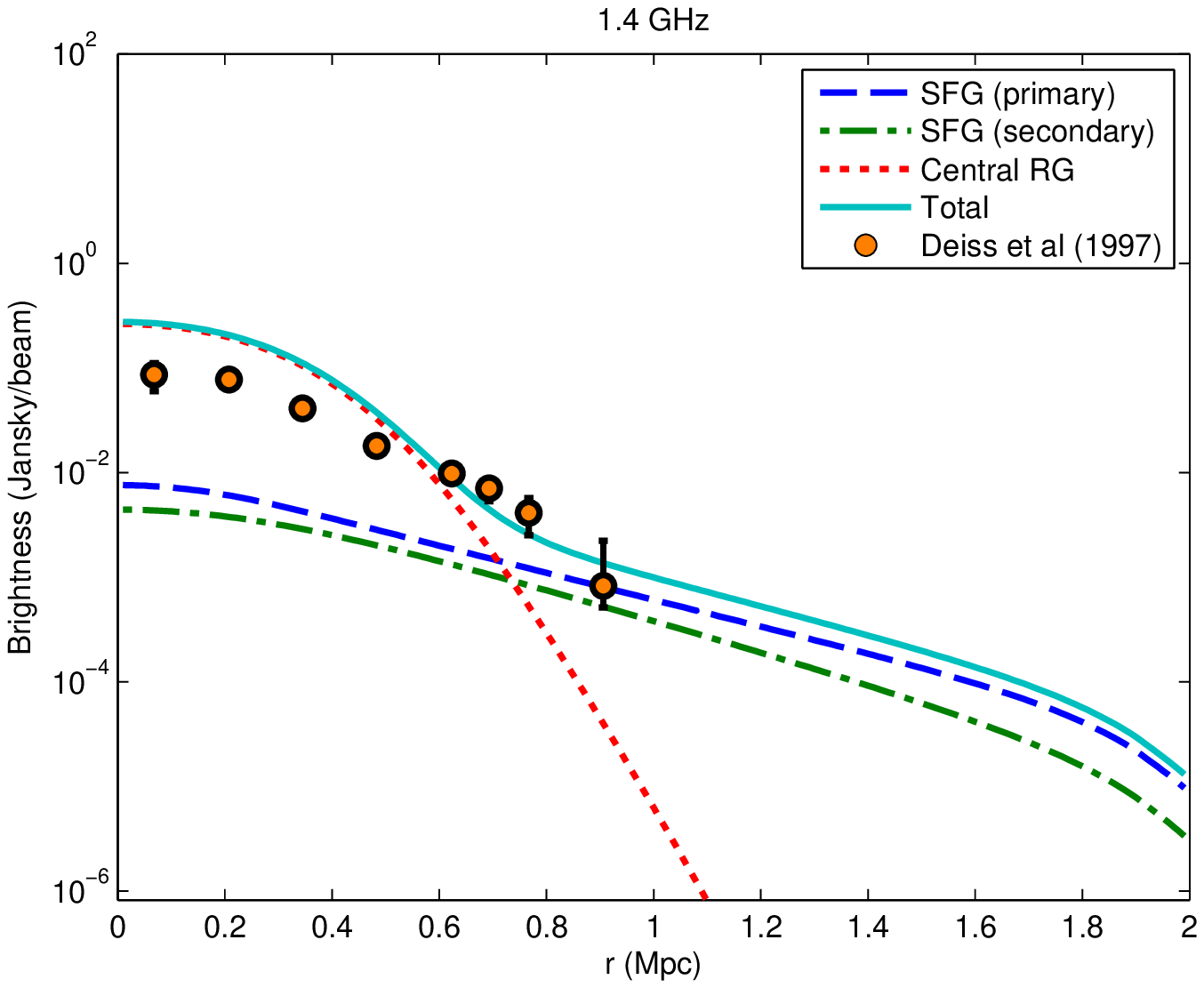,width=5.7cm,height=4.5cm}
\caption{Left panel: the total spectral radio luminosity of primary and secondary electrons 
from SFGs, and electrons from central radio galaxies 
(solid curve); also shown are observational data compiled in Ref. ~[\onlinecite{p10}]. 
Middle panel: measured and predicted profiles of the radio brightness at 350 MHz. 
Dashed, dashed-dotted, and dotted curves show the profiles of radio emission by primary 
and secondary electrons from SFGs, and by electrons from central radio galaxies, 
respectively; the solid curve is the sum of all three contributions. The circles with 
error bars represent the Brown \& Rudnick (2011, Ref. [\onlinecite{br11}]) data in the 
region $r\simeq 0.04-1.4$ Mpc. {\it It is important to realize that the 350 MHz measurements 
at $r > 0.7$ Mpc include appreciable emission from the relic Coma A which is \bf{not included} 
in our model}. Right panel: measured and predicted profiles of the radio brightness at 1.4 GHz. 
Model curves are as in the middle panel; data points are based on measurements of halo emission 
by Deiss \ea (1997) (Ref. [\onlinecite{d97}]).}
\label{fig:syspds}
\end{figure*}

Energetic protons interacting with ambient IC nuclei produce both charged and neutral pions; 
the former produce secondary electrons and positrons whose spectral density can be derived from 
the spectral density of energetic protons using [\onlinecite{d86}]. 
\begin{equation}
\begin{split}
f_{\pi}(E_{\pi})={}&\frac{2}{3}\int_{E_{th}}^{\infty}f_p(E_p)f_{\pi}(E_{\pi},E_p)dE_p, 
\label{eq:pip}
\end{split}
\end{equation}
where E$_p$ is the (total) proton energy in GeV, E$_{th}$=1.22 GeV is the 
threshold for pion creation, $f_p(E_p)$ is the proton distribution, and 
$f_{\pi}(E_{\pi},E_p)=4cn_p(r)\xi(E_p)\sppi(E_p)\delta(E_p-4{E}_{\pi})$ is the pion distribution. 
Here, $n_p(r)$ is the target nucleon density, $\xi(E_p)\approx 2$ is the hadronic reaction multiplicity, 
and $\sppi(E_p)$ is the cross section for pion production, calculated using an analytical fit [\onlinecite{kab06}]. Substituting $f_{\pi}(E_{\pi},E_p)$ in  Eq.~(\ref{eq:pip}), then yields 
$f_{\pi}(r,E_{\pi})=16cn_p(r)\sppi(4E_{\pi}) f_{p}(4E_{\pi})$. For a proton escape rate 
$q_{p}(\gp)=q_{0,p}\gp^{-\delta_{p}}$, with spectral index $\delta_{p}$, the pion distribution 
expressed in terms of the pion Lorentz factor becomes $f_{\pi}(r,\gamma_{\pi})=
0.74N_{0,p}cn_p(r)\sppi(4E_{\pi}) \cdot (0.56\gamma_{\pi})^{-\delta_{p}}$. 
Note that the accuracy of the above treatment can be improved by using the more detailed 
calculations of pion yields in p-p interactions in Ref. [\onlinecite{kmo14}]; given the 
considerable uncertainties basic parameter values in our generic model, use of the approximate 
relations in Ref. [\onlinecite{d86}] is adequate for our purposes here.

The distribution function of electrons produced by charged pion decay can be approximated [\onlinecite{rl66}] as $f_e(r,\gamma_e)= f_{\pi}(r,\gamma_{e}/A)/A$, where 
$A=(1/4)(m_{\pi}/m_e)\approx 70$. Calculation of the secondary electron distribution is 
carried out by tracking the protons diffusing out from their sources to the point of electron creation, 
yielding proton distributions analogous to Eq.~(\ref{eq:fgd}), and consequent diffusion of secondary 
electrons to the observation point. Using the expression for $f_e(r,\gamma_e)$ to transform the 
distribution of the pions into that of pre-diffusion electrons as 
$f_e^{*}(\tm,\gte,\rmm)=f_{\pi}^{*}(\tm,\gte/A,\rmm)/A$, and substituting the electron escape 
rate in Eq.~(\ref{eq:fgd}), then yields the distribution of secondary electrons.

\subsection{Energy Loss Rates \& Magnetic Field Profile}

Energy loss processes for IC energetic electrons are electronic (Coulomb) excitations, 
bremsstrahlung, and Compton-synchrotron, with the total energy loss rate given by 
\be
b(\gamma)=b_{0} + b_{1}\gamma + b_{2} \gamma^{2}  \, ,
\ee
where $b_{0} \simeq 1.1\times 10^{-15}(n_{e}/10^{-3}cm^{-3}) \,s^{-1}$ is the loss 
rate by electronic excitations in ionized gas [\onlinecite{g72}]; $n_{e}$ is the  IC electron density. 
The coefficients of the bremsstrahlung and Compton-synchrotron rates are [\onlinecite{bg70}] 
$b_{1} \simeq 1.3\times 10^{-18}\times (n_{e}/10^{-3}cm^{-3}) \,s^{-1}$, and 
$b_2=1.3\times 10^{-20}[(1+z)^{4}+0.1\times (B/(10^{-6}\mu G))^2] \,\,s^{-1}$, respectively. 
Compton scattering of the electrons is by the CMB, whose energy density increases with redshift, 
$z$. The main energy loss of energetic protons is by p-p interactions at a rate 
$b(\gp)=n_p(r)\,c\,\kappa\,\sinel\gp$, where $\kappa\approx 0.45$ and $\sinel$ are the mean 
inelasticity and cross section in inelastic p-p interactions, respectively.
(Note that the above expressions are valid for $\gamma \gg 1$; the correct expressions at the 
trans-relativistic regime can be found in [\onlinecite{rp15}].)  

A realistic profile of the mean IC magnetic field is of central importance for reliable calculations 
of the level and spatial distribution of the halo emission. In the hot, fully-ionized, highly electrically conductive IC gas, magnetic fields are likely frozen into the plasma; if so, the field spatial 
dependence is expected to scale as $n_{e}^{2/3}$ [\onlinecite{r88}]. The index in this 
scaling is somewhat lower, $1/2$, if magnetic energy density is in equipartition with the gas 
energy density.

\section{Results}

Electron and proton escape rates from the Galaxy were deduced based on results obtained 
by Strong \ea (2010; Ref. [\onlinecite{sea10}] ) from a set of diffusion and diffusive 
re-acceleration models with different Galactic halo sizes. From the range of values for their 
different models we determine average rates, $q_{i}(\gamma)=q_{i,0}\gamma^{-\delta_{i}}$, 
that are $q_{e,0}\simeq 5\times10^{45}$ s$^{-1}$, $\delta_{e}=2.5$ for electrons, and 
$q_{p,0}\simeq 2.1\times10^{43}$ s$^{-1}$ and $\delta_{p}=2.4$ for protons. As noted, 
we take the number of Galaxy-equivalent SFGs to be $10^2$.

In a 1.4 GHz survey of Coma [\onlinecite{mea09}], 19 elliptical radio galaxies were 
identified as definite cluster members, 
including the very strong NGC4869 \& NGC4874 in the cluster core. The emission of the 
other 17 elliptical radio is dominated by the NGC4839 and NGC4827, which are part of the 
infalling NGC4839 group in the SW region of the cluster, the radio relic Coma A. We confine 
our modeling to the nearly spherical radio halo, and since the extended emission of Coma A is 
outside the central $\sim 1$ Mpc radial region, we ignore the contribution of all the 17 galaxies 
to the electron density. Radio emission scales as $N_{e }B^{1+\alpha}$, where $N_{e}$ is the 
total number of emitting electrons and $\alpha$ is the radio power-law index, but since the estimated 
mean magnetic fields in the extended radio emitting regions of the two central radio galaxies 
([\onlinecite{fg87}], \onlinecite{fea90}]) are about the same as the mean Galactic field ($B 
\sim 6\,\mu$G), we scale the Galactic electron rate by the ratio of the combined radio luminosity 
of these galaxies to that of the Galaxy. Doing so yields a spectral escape rate of 
$3\times 10^{48}$ s$^{-1}$, taking $\delta_{e}=2.5$; in our calculations we adopt a value 
that is lower by a factor of 2/3.

The distributions of electrons and protons escaping SFGs and their yields are evolved for $6$ 
Gyr (over which the {\it electron} distribution is essentially in its asymptotic steady-state), 
whereas for electrons from the central radio galaxies we show results for $1$ Gyr, a more 
typical time for their AGN-powered emission, either continuously or successively. This period 
of activity is reasonable given the long duration of mergers and strong tidal interactions, 
processes that enhance particle acceleration in galactic nuclei. Other key parameters in the 
calculation are the central value of the cluster magnetic field, $B_0 =6\,\mu$G (somewhat 
lower than deduced from Faraday Rotation measurements [\onlinecite{fea95}]).
 
A wide range of magnetic field coherence scales is expected, reflecting the varying properties 
of IC plasma, with values as low as a few tenths of a kpc to $\sim 10$ kpc. A fiducial value 
of $1$ kpc is adopted, for which $D=3\times 10^{31}$ cm$^2$ s$^{-1}$.

Having selected Coma ($z=0.023$) as our fiducial cluster, we adopt the ($\beta$-like) gas 
density profile used in the analysis of {\it Planck} SZ measurements [\onlinecite{ade13}] 
\begin{equation}
n^2(r)=\frac{n_{0}^{2}}{\left[1+(r/r_c)^2\right]^{3\beta}}
\frac{1}{\left[1+(r/r_s)^3\right]^{\epsilon/3}}
\end{equation}
with $n_0=2.9\times 10^{-3}$ cm$^{-3}$, $r_{c}\simeq0.4$ Mpc,
$\beta \simeq0.57$ $r_{s}\simeq 0.7$ Mpc, and $\epsilon\simeq1.3$. 

We calculated the spectral radio synchrotron and Compton (in the Thomson limit) X-ray and 
$\gamma$-emissivities by relativistic electrons, traversing IC fields and scattering off the 
CMB, using the relevant expressions in Ref. [\onlinecite{bg70}]. The spectral $\gamma$-ray 
emissivity from neutral pion decay was calculated using the expression in Ref. [\onlinecite{ms94}].

The spectral and spatial distributions of radio emission by primary and secondary electrons 
originating in SFGs, and the corresponding emission by primary electrons from central 
radio galaxies, are shown in Figure~(\ref{fig:syspds}). Our fiducial spectral profile seems to be 
somewhat flatter than the trend seen in the data above $\sim 1$ GHz. In fact, a linear fit to current 
data yields a spectral index of $1.4\pm 0.2$  (at 95\% significance) that implies a somewhat 
steeper spectral profile of electrons emerging from radio and SFGs, i.e., $\delta_{e} \simeq 2.8$, 
rather than the value of $2.5$ adopted in our generic model. 
Also shown are measurements of the Coma halo compiled in Ref. [\onlinecite{p10}]. 

For a direct comparison with the profile deduced from the 350 
MHz measurements reported in Ref. [\onlinecite{br11}], and plotted in Figure 10 in 
Ref. [\onlinecite{ade13}], we convolved the (line of sight) integrated 
emissivities with a gaussian 10' (FWHM) beam. We note that the plotted 350 MHz data were 
obtained by azimuthally averaging the substantially angle-dependent emission profile (shown 
in Ref. [\onlinecite{ade13}]). Above $\sim 0.8$ Mpc, the plotted profile includes emission from 
the relic Coma A, which is locally as much as a factor $\sim 6$ higher than the emission in the 
SE part of the cluster. Subtracting the flux of Coma A in the the SW 0.8-1.4 Mpc region would 
bring the data points appreciably closer to our profile. Indeed, the profile of the Coma halo was 
measured also at 1.4 GHz smoothed with
9.35' beam) by Deiss \ea (1997, Ref. [\onlinecite{d97}]), whose data cover only the main halo 
region, $r\leq 1$ Mpc. As is obvious from the right panel of Figure~(\ref{fig:syspds}), our profile 
agrees well with this steeper profile that does not include substantial emission from Coma A. 
Comparison of the measured profiles demonstrates the increased significance of the relative 
contribution of emission by electrons from SFGs, clearly establishing the need for SFGs as 
electron sources.

\begin{figure}[t]
\centering
\epsfig{file=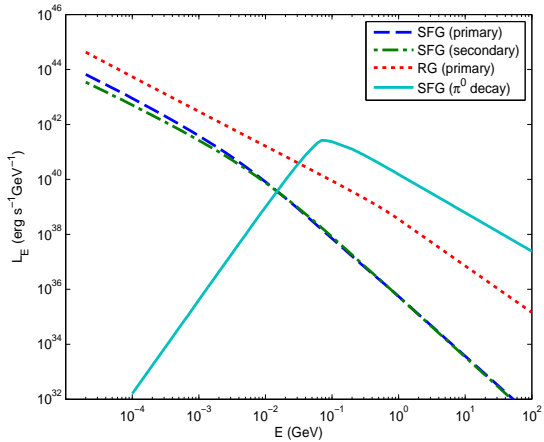,width=8.0cm,height=6.3cm,clip=}
\caption{Spectral \gr luminosity from neutral pion decay and Compton scattering of 
relativistic electrons off the CMB.}
\label{fig:gampd}
\end{figure}

The spatially integrated 350 MHz flux from Coma is somewhat lower than that deduced 
from the measured profile, a difference that is reflected in the respective spatial profiles 
in Figure~(\ref{fig:syspds}), with the predicted emission more pronounced in the cluster core, 
but this emission falls steeply and cannot account for the measured emission beyond $\sim 
0.5$ Mpc. We emphasize that our aim here is to present generic features of our approach, 
rather than best-fitting model parameters to the specific Coma spectral and spatial measurements. 
Therefore, the fact that the predicted emission does not exactly match the observations is actually 
not surprising given the appreciable range of parameter  ($D$, $B$, $q_{i}$) values. 

Spectral $\gamma$-ray luminosity by neutral pion decay is shown in Figure~(\ref{fig:gampd}). 
The emission is spatially extended reflecting the large diffusion radii of the parent protons. 
Also shown in the figure are the spectral distributions of X-and-$\gamma$-ray emission by 
Compton scattering of relativistic electrons off the CMB, whose combined power-law spectral 
luminosity has a mean index of $\sim 1.3$. The emission is clearly more extended than in the 
radio (due to the CMB spatial uniformity). The integrated luminosity above 100 MeV, 
$\sim 1\times 10^{41}$ erg s$^{-1}$, is mostly from $\pi^{0}$ decay.

\section{Discussion}

The approach adopted in our work is minimalistic as it rests on known properties of galactic 
energetic particles, quantitative description of their diffusion in IC space and consequent radiative 
yields, thereby establishing the level of radio emission beyond which turbulent (re)acceleration might 
be needed to fully account for halo emission, if electrons and protons from star-forming and other 
widely distributed galaxies cannot account for the relatively high 350 MHz emission measured in the 
outer region of the Coma cluster. A basic result of our work is that with either known or estimated 
values of these quantities, the predicted level of the radio luminosity can be comparable to the 
measured value. The significance of this result rests also on the fact that we have not included any 
additional sources of energetic particles, such as AGN or radio galaxies other than NGC 4869 \& 
NGC 4874.

Of primary importance are the proton and electron escape fractions from a SFG, which are based 
on the detailed study in Ref. [\onlinecite{sea10}]. Whereas for protons this fraction is only a few 
percent, it is about half for electrons; if anything, the calorimetric fraction is expected to be lower 
due to reduced halo size of a SFG in the more tidally interactive cluster environment.
\footnote{Evidence for a relatively extended radio halo of SFGs has recently been obtained 
[\onlinecite{wie15}] from a stacking analysis of 1.5 GHz measurements of 30 nearby field galaxies.} 
Nonetheless, the exact values of the electron escape rates from radio galaxies are uncertain; a 
lower escape rate than estimated here would reduce the level of radio emission in the core, 
weakening the discrepancy with the measured 1.4 GHz profile. The dependence of synchrotron 
emission on the magnetic field, $B^{\alpha +1}$ ($\sim B^{2.3}$ for Coma) is sufficiently steep 
to make the feasibility of detection of a low brightness radio halo very sensitive to specific 
conditions in the cluster. Our fiducial {\it central} value, $B_{0}= 6\, \mu$G, implies a 
volume-averaged value of only $0.3\, \mu$G across the cluster. 

Extended radio emission is far less ubiquitous than X-ray emission from IC gas. A major issue 
then is what cluster evolutionary processes give rise to this emission; suggestions abound, 
ranging from strong radio galaxies and AGN as sources of energetic particles, and particle 
acceleration by turbulent shocks excited by galactic motions or by merger shocks 
(e.g., Ref. [\onlinecite{bj14}] ). Various scaling relations reflecting correlations between radio 
emission and cluster properties have been either deduced or invoked in support of specific models. 
However, such relations may not always be reliable due to inherent parameter degeneracies, 
reflecting the considerable coupling between respective cluster properties and evolutionary 
processes. An example for this inherent limitation in testing a specific model for the origin of 
extended radio emission is the commonly advocated view that energetic particles are accelerated 
by merger shocks. Indeed, during cluster hierarchical evolution, mergers of sub-clusters can 
create conditions under which shocks may have the properties required for efficient particle 
acceleration. But mergers enhance star formation in cluster galaxies, which also results in 
increased SN activity with the same end result of particle acceleration; thus, a basic evolutionary 
process can be generically the root cause of different models for the origin of enhanced particle 
acceleration. 

The predicted level of \gr emission from energetic protons and electrons emanating from SFGs 
and radio galaxies in a Coma-like cluster is $L (>100 MeV)=O (10^{41})$ erg\,s$^{-1}$. 
A recent search for \gr emission from the Coma region in the {\it Fermi}/LAT 6 yr database 
has not yielded a statistically significant detection [\onlinecite{ack16}] ; the implied upper limit 
on emission above $100$ MeV is about an order of magnitude higher than the level predicted 
here. Obviously, this is a bound on the total emission from the Coma region, including the intrinsic 
emission from all radio galaxies and SFGs. Multiplying our overall normalization factor ($N_{s}$) 
by the total \gr luminosity of the Galaxy [\onlinecite{sea10}] we obtain $\sim 7\times 10^{40}-1\times10^{41}\,erg\,s^{-1}$ for the range of models considered in the latter paper. 

Also of much interest is the predicted NT X-ray luminosity due to Compton scattering of 
relativistic electrons off the CMB. As shown in the right panel of Figure 2, most of the Compton 
emission is predicted to be by primary electrons originating in the central radio galaxies; the 
total (including contributions of electrons from SFGs) $0.02-1$ MeV luminosity is $L \sim 
2.7\times 10^{40}$ erg\,s$^{-1}$. Dedicated searches for NT emission above 20 keV began 
long ago [\onlinecite{rgr87}] and continued with almost all successive X-ray satellites. Results 
of all these searches (e.g., Refs. [\onlinecite{r08}], [\onlinecite{wik14}]) yielded no unequivocal 
detections. For Coma, extensive RXTE measurements [\onlinecite{rg02}] yielded the upper limit 
$L (20-80 \,keV) \leq 3\times 10^{42}$ erg\,s$^{-1}$, much higher than the luminosity of IC 
electrons predicted here. However, our estimated level is clearly a lower limit on the total cluster 
emission since it does not include intrinsic emission from galactic sources.

\begin{acknowledgements}
We are grateful to the referee for constructive evaluation of our work and for several useful 
suggestions that improved presentation, and to Dr. Larry Rudnick for a useful discussion. 
This work has been supported by a grant from the Israel Science Foundation. 
\end{acknowledgements}

\end{document}